# High-temperature dependence of anomalous Ettingshausen effect in SmCo$_5$-type permanent magnets

Asuka Miura[1], Keisuke Masuda[1], Takamasa Hirai[1], Ryo Iguchi[1], Takeshi Seki[1,2,3], Yoshio Miura[1,4], Hiroki Tsuchiura[3,5], Koki Takanashi[2,3,6], and Ken-ichi Uchida[1,2,3,a]

[1] National Institute for Materials Science, Tsukuba 305-0047, Japan.
[2] Institute for Materials Research, Tohoku University, Sendai 980-8577, Japan.
[3] Center for Spintronics Research Network, Tohoku University, Sendai 980-8577, Japan.
[4] Center for Spintronics Research Network, Osaka University, Osaka 560-8531, Japan.
[5] Department of Applied Physics, Tohoku University, Sendai 980-8579, Japan.
[6] Center for Science and Innovation in Spintronics, Core Research Cluster, Tohoku University, Sendai 980-8577, Japan.
[a] E-mail: UCHIDA.Kenichi@nims.go.jp

**ABSTRACT**
The anomalous Ettingshausen effect (AEE) in SmCo$_5$-type permanent magnets has been investigated in the high-temperature range from room temperature to around 600 K. The anomalous Ettingshausen coefficient of the SmCo$_5$ and (SmGd)Co$_5$ magnets monotonically increases with increasing the temperature and shows the similar temperature dependence, while the coefficient of SmCo$_5$ is slightly larger than that of (SmGd)Co$_5$ at high temperatures. The dimensionless figure of merit for AEE in SmCo$_5$ at high temperatures is much greater than the previous record obtained for the anomalous Nernst effect. The observed high-temperature behavior of AEE is discussed based on the first-principles calculations of transverse transport coefficients.

The anomalous Ettingshausen effect (AEE) refers to the conversion of a charge current into a transverse heat current in a magnetic conductor. The AEE-induced heat current density is described as

$$\mathbf{j}_{q,\text{AEE}} = \Pi_{\text{AEE}} \left( \mathbf{j}_c \times \mathbf{m} \right), \tag{1}$$

where $\mathbf{j}_c$, $\mathbf{m}$, and $\Pi_{\text{AEE}}$ denote the charge current density, unit vector of magnetization, and anomalous Ettingshausen coefficient, respectively [Fig. 1(a)].[1-10] Since AEE works as temperature modulators with simple structure and versatile scaling owing to the unique thermoelectric conversion symmetry, it may pave the way for thermal management technologies for electronic and spintronic devices. Experimental studies on AEE have been developed rapidly since the establishment of the versatile measurement technique for this phenomenon.[4]

Recently, large AEE was observed in SmCo$_5$-type permanent magnets at room temperature.[8] This result indicates that SmCo$_5$-type magnets also exhibit good performance for the thermoelectric generation based on the anomalous Nernst effect (ANE),[11-26] which is the Onsager reciprocal of AEE. The figure of merit for AEE/ANE in SmCo$_5$-type magnets at room temperature is comparable to the record-high value in a Heusler ferromagnet.[19] In addition to the thermoelectric conversion properties, permanent magnets have several practical advantages over soft magnetic materials for AEE/ANE applications. First, in permanent magnets, AEE/ANE works in the absence of magnetic fields and its performance is not affected by external field disturbance, where static magnetic properties of the magnets are important to determine the zero-field performance of AEE/ANE devices. Second, permanent magnets are mass-produced and widely distributed in a society, enabling the construction of ubiquitous, low-cost, and large-area AEE/ANE devices. Therefore, clarifying the thermoelectric conversion performance of AEE/ANE in $SmCo_5$-type magnets is important.

In this study, we report the observation of AEE in $SmCo_5$-type magnets in the high-temperature range from room temperature to around $T = 600$ K. We focus on this temperature range because of the working temperature limit of $SmCo_5$-type magnets; as shown in Fig. 1(b), the remanent magnetization $M_r$ of the $SmCo_5$-type magnets sharply decreases around 700 K. The previous study shows that, although the $SmCo_5$-type magnets with different Gd contents have substantially different static magnetic properties and microstructures, $\Pi_{AEE}$ is almost independent of the composition at room temperature.[8] However, the behaviors of AEE at high temperatures are yet to be clarified.

The samples used in this study are polycrystalline $SmCo_5$ and $(SmGd)Co_5$ slabs with a rectangular cuboid shape, commercially available from Magfine Corporation, Japan. The Sm:Gd ratio in the $(SmGd)Co_5$ slab is 48:52 in at%. To estimate the figure of merit for AEE and the Seebeck effect in the $SmCo_5$ and $(SmGd)Co_5$ slabs, we measured the $T$ dependence of $\Pi_{AEE}$, the longitudinal electrical conductivity $\sigma_{xx}$, thermal conductivity $\kappa$, and Seebeck coefficient $S_{xx}$. The sample size for the measurements of $\Pi_{AEE}$, $\sigma_{xx}$, and $S_{xx}$ ($\kappa$) is 10.0 or 12.0 × 1.5 × 1.5 mm³ (10.0 × 10.0 × 2.0 mm³), where the samples are magnetized along the 1.5 mm (10.0 mm) direction in the absence of magnetic fields. AEE was measured by means of the lock-in thermography (LIT) technique,[4-10,27-29] as detailed below. The $T$ dependence of $\sigma_{xx}$ and $S_{xx}$ was simultaneously measured with the Seebeck coefficient/electric resistance measurement system (ZEM-3, ADVANCE RIKO, Inc). The $T$ dependence of $\kappa$ was determined through thermal diffusivity measured by the laser flash method, specific heat measured by the differential scanning calorimetry, and density measured by the Archimedes method.

The procedures for the AEE measurements are as follows. The LIT method allows us to estimate $\Pi_{AEE}$ of permanent magnets without applying external magnetic fields, when $M_r$ is finite.[8] We measured thermal images of the surface of the $SmCo_5$ or $(SmGd)Co_5$ slab with an infrared camera while applying a square-wave-modulated AC charge current with the frequency $f$, amplitude $J_c$ (with the density $j_c$), and zero offset to the slab in the $x$ direction and extracted the first harmonic response of the detected images, which are transformed into the lock-in amplitude $A$ and phase $\phi$ images through Fourier analysis [Fig. 1(f)]. Based on this procedure, thermoelectric signals ($\propto J_c$) free from Joule-heating signals ($\propto J_c^2$) can be detected.[4-10] The thermoelectric signals around the center of the sample are due purely to AEE, which are separated from parasitic signals due to the Peltier effect generated at the ends of the sample. The sample was fixed on a sapphire substrate attached to the top of a Cu stage of which the temperature can be tuned with a heater and a resistance temperature sensor embedded in the stage [Fig. 1(f)]. To enhance the infrared emissivity, the top surface of the sample was coated with insulating black ink. The LIT measurements were performed in a high vacuum and thermal images were monitored through an infrared-transparent $CaF_2$ window. The effect of the $CaF_2$ window to the thermal images was calibrated by comparing the infrared intensity on the surface of a reference thermometer, measured with the infrared camera through the window, and the output of the thermometer. This calibration allows us to monitor the sample temperature during the LIT measurements through steady-state thermal images [Fig. 1(e)]. Figures 1(c) and 1(d) respectively show an example of the $A$ and $\phi$ images for the $SmCo_5$ slab at $T = 317$ K, where the remanent magnetization (charge current) is along the $z$ ($x$) direction. Clear current-induced temperature modulation was observed to appear on the entire surface of the slab. This behavior is consistent with the symmetry of AEE; the temperature modulation signal is generated as a result of the AEE-induced heat current along the $y$ direction [Fig. 1(a)].

Figure 2(a) shows the $T$ dependence of $A/j_c$ for the SmCo$_5$ and (SmGd)Co$_5$ slabs at $f = 10$ Hz, where the data points were obtained by averaging the temperature modulation signals on the area defined by the white rectangle with a size of 100 × 40 pixels in Fig. 1(e). The AEE-induced temperature modulation appears in both the slabs in the whole temperature range studied without changing its sign [see the inset to Fig. 2(a)].

For the quantitative discussion, we estimated the magnitude of the AEE-induced temperature modulation in the steady state, *i.e.*, at $f = 0$ Hz, by fitting the $f$ dependence of the AEE signal with the solution of the heat equation in the frequency domain [see Fig. 2(b) and Ref. 8 for details]. From the steady-state AEE signal, we can determine the anomalous Ettingshausen coefficient as $\Pi_{\text{AEE}} = \kappa \Delta T / \tilde{j}_c L$, where $\tilde{j}_c$, $\Delta T$, and $L$ are the sinusoidal amplitude of the charge current density, temperature difference between the top and bottom surfaces of the slab, and thickness of the slab (= 1.5 mm), respectively (note that $\Pi_{\text{AEE}}$ is independent of $L$ since $\Delta T$ is proportional to $L$).[7] Here, recall the fact that the magnetization of the SmCo$_5$ and (SmGd)Co$_5$ slabs was not fully saturated during the AEE measurements [Fig. 1(b)]. We thus set $\Delta T = 2A_{0\text{Hz}}(M_s/M_r)$ to take the $T$-dependent difference between $M_r$ and $M_s$ into consideration, where $A_{0\text{Hz}}$ and $M_s$ denote $A$ at $f = 0$ Hz and the saturation magnetization, respectively. We can obtain the anomalous Nernst coefficient $S_{\text{ANE}}$ via the Onsager reciprocal relation: $\Pi_{\text{AEE}} = S_{\text{ANE}}T$.

In Fig. 3(a), we show the $T$ dependence of $\Pi_{\text{AEE}}$ for SmCo$_5$ and (SmGd)Co$_5$. We found that $\Pi_{\text{AEE}}$ monotonically increases with increasing $T$ for both the slabs and the $T$ dependence of $\Pi_{\text{AEE}}$ is quite similar to each other, while $\Pi_{\text{AEE}}$ of SmCo$_5$ is slightly larger than that of (SmGd)Co$_5$ at high temperatures. The weak dependence of AEE on the Gd content in the SmCo$_5$-type magnets is consistent with the previous result.[8] The corresponding $S_{\text{ANE}}$ values are shown in Fig. 3(b); $S_{\text{ANE}}$ also monotonically increases with $T$ but its $T$ dependence is much weaker than that of $\Pi_{\text{AEE}}$ due to the factor $T$ in the Onsager reciprocal relation. The maximum $\Pi_{\text{AEE}}$ ($S_{\text{ANE}}$) value for SmCo$_5$ is $3.2 \times 10^{-3}$ V ($5.6 \times 10^{-6}$ VK$^{-1}$) at $T = 572$ K. We confirmed negative (no) correlation between $S_{\text{ANE}}$ and $M_s$ for SmCo$_5$ [(SmGd)Co$_5$], which clearly deviates from the scaling behavior reported in Ref. 17 [see the inset to Fig. 3(b)].

Figure 3(c) shows the $T$ dependence of the dimensionless figure of merit for AEE/ANE in the SmCo$_5$ and (SmGd)Co$_5$ slabs, which was estimated based on the following definition:[8,30]

$$Z_{\text{AEE}}T = \frac{\Pi_{\text{AEE}}^2 \sigma_{xx}}{\kappa} \frac{1}{T} \quad \left( = \frac{S_{\text{ANE}}^2 \sigma_{xx}}{\kappa} T \right). \quad (2)$$

By using the $T$ dependence of $\Pi_{\text{AEE}}$ [Fig. 3(a)] together with $\sigma_{xx}$ and $\kappa$ [Fig. 3(d)], we obtained the $Z_{\text{AEE}}T$ values. Although the $T$-dependent increase of $\Pi_{\text{AEE}}$ is partially offset by the $T$-dependent decrease of $\sigma_{xx}$ and the factor $1/T$ in Eq. (2), the resulting $Z_{\text{AEE}}T$ monotonically increases with $T$ for both the slabs. The estimated $Z_{\text{AEE}}T$ value for SmCo$_5$ is $0.5 \times 10^{-3}$ ($1.1 \times 10^{-3}$) at $T = 317$ K (572 K). Significantly, the $Z_{\text{AEE}}T$ value at the high temperature marks a record high for AEE/ANE, which is much higher than the value of full-Heusler Co$_2$MnGa.[19] We also note that even the maximum $Z_{\text{AEE}}T$ value for SmCo$_5$ is one order of magnitude smaller than the figure of merit for the Seebeck effect $Z_{\text{SE}}T$ [Figs. 3(c) and 3(f)], which is due to the fact that $|S_{\text{ANE}}|$ is several times smaller than $|S_{xx}|$ [Figs. 3(b) and 3(e)].

We are now in a position to discuss the observed $T$ dependence of AEE/ANE in SmCo$_5$. The following discussion is mainly presented in terms of ANE because of the simple notation. In general, the anomalous Nernst coefficient can be divided into the two terms:[24]

$$S_{\text{ANE}} = \rho_{xx}\alpha_{xy} + \rho_{xy}\alpha_{xx} \equiv S_\text{I} + S_\text{II}, \qquad (3)$$

where $\rho_{xx} = 1/\sigma_{xx}$ ($\rho_{xy} = -\sigma_{xy}/\sigma_{xx}^2$) denotes the diagonal (off-diagonal) component of the electrical resistivity tensor, $\alpha_{xx}$ ($\alpha_{xy}$) the diagonal (off-diagonal) component of the thermoelectric conductivity tensor, $S_\text{I} = \rho_{xx}\alpha_{xy} = \alpha_{xy}/\sigma_{xx}$, and $S_\text{II} = \rho_{xy}\alpha_{xx} = -S_{xx}\sigma_{xy}/\sigma_{xx}$. The $S_\text{I}$ term is an intrinsic part of ANE as it originates from the transverse thermoelectric conductivity $\alpha_{xy}$, which is determined by the energy derivative of the anomalous Hall conductivity $\sigma_{xy}$. In contrast, the $S_\text{II}$ term is attributed to the concerted action of the Seebeck and anomalous Hall effects. In Ref. 8, we showed that AEE/ANE in the SmCo$_5$-type magnets at room temperature originates mainly from the $S_\text{I}$ term owing to the large $\alpha_{xy}$ and the experimental results are well reproduced by the first-principles calculations of the transverse transport coefficients. To investigate the $S_\text{I}$ and $S_\text{II}$ contributions at high temperatures, we compare the experimental results with the first-principles calculations based on the same procedures shown in Ref. 8. Here, we calculated $\sigma_{xy}$ and $\alpha_{xy}$ of SmCo$_5$ with changing $T$ and the chemical potential $\mu$, where the electronic band structure at absolute zero temperature is assumed and the $T$-dependent effects are included through the Fermi distribution function. By combining the calculated $\alpha_{xy}$ ($\sigma_{xy}$) values with the observed $\sigma_{xx}$ ($S_{xx}$ and $\sigma_{xx}$) values for SmCo$_5$, we estimated the $S_\text{I}$ ($S_\text{II}$) contribution.

Figures 4(a) and 4(b) show the $\mu$ dependence of $\sigma_{xy}$ and $\alpha_{xy}$ for SmCo$_5$ at various temperatures, respectively. At $\mu = 0$ eV, the calculated $\alpha_{xy}$ shows large values of $> 6$ Am$^{-1}$K$^{-1}$ and monotonically increases with increasing $T$, while $\sigma_{xy}$ is very small above room temperature. As shown in Fig. 4(c), the calculated $T$ dependence of $S_{\text{ANE}}$ at $\mu = 0$ eV is dominated by the $S_\text{I}$ contribution and the $S_\text{II}$ values are negligibly small. In this case, the calculation cannot explain the high-temperature behavior of AEE/ANE for SmCo$_5$; the difference between the observed and calculated $S_{\text{ANE}}$ values increases with $T$. The discrepancy between ANE experiments and first-principles calculations often occurs, but it has been reported that the experimental results can be explained by considering the $\mu$ shift due to carrier doping.[21,26] We thus adopted a similar strategy; by comparing the experimental and calculated results with changing $\mu$, we found that the calculation at around $\mu = -0.14$ eV well reproduces the observed $T$ dependence of $S_{\text{ANE}}$ for SmCo$_5$ over the $T$ range of interest, where the $S_\text{I}$ ($S_\text{II}$) contribution at $\mu = -0.14$ eV is smaller (larger) than that at $\mu = 0$ eV [Fig. 4(c)]. The calculated ratio between $S_\text{I}$ and $S_\text{II}$ at $\mu = -0.14$ eV is quantitatively consistent with the experimental results at room temperature.[8] These results suggest that the high-temperature behavior of AEE/ANE for SmCo$_5$ is attributed to its intrinsic band structure and $\mu$ shift, where the hole doping of $\mu = -0.14$ eV corresponds to a change in the number of valence electrons of 1.32 per a SmCo$_5$ unit cell. We also note that the weak dependence of AEE on the Gd content can be explained by the fact that the electronic structure of the SmCo$_5$-type magnets near the Fermi energy is determined mainly by the Co sublattice.[31] To confirm the validity of this interpretation and clarify the carrier-doping effects, systematic investigations on elemental substitution and impurity effects on the Co sublattice in the SmCo$_5$-type magnets and direct observation of their band structures are necessary.

In conclusion, we have investigated the transport properties in the SmCo$_5$-type permanent magnets and observed large AEE above room temperature. The anomalous Ettingshausen coefficient of the SmCo$_5$ and (SmGd)Co$_5$ magnets monotonically increases with increasing the temperature $T$. Both the magnets have comparable $T$-dependent AEE coefficients, while the magnitude of the AEE signals for SmCo$_5$ is slightly larger than that for (SmGd)Co$_5$ at high temperatures. The dimensionless figure of merit for AEE in SmCo$_5$ reaches ~ $1 \times 10^{-3}$ when $T > 500$ K, which is the maximum value reported so far. The $T$ dependence of AEE in SmCo$_5$ can be explained by the first-principles calculations typically used for estimating the intrinsic contribution of AEE/ANE if the chemical potential shift is taken into account. These results provide a crucial piece of information for understanding the microscopic mechanism of AEE/ANE in the SmCo$_5$-type magnets and for creating thermoelectric converters based on permanent magnets.

The authors thank Y. Sakuraba for valuable discussions. This work was supported by CREST "Creation of Innovative Core Technologies for Nano-enabled Thermal Management" (JPMJCR17I1) from JST, Japan, Grant-in-Aid for Scientific Research (S) (JP18H05246) and Grant-in-Aid for Scientific Research (B) (JP19H02585) from JSPS KAKENHI, Japan, and the Canon Foundation. A.M. and T.H. are supported by JSPS through Research Fellowship for Young Scientists (JP18J02115, JP20J00365).

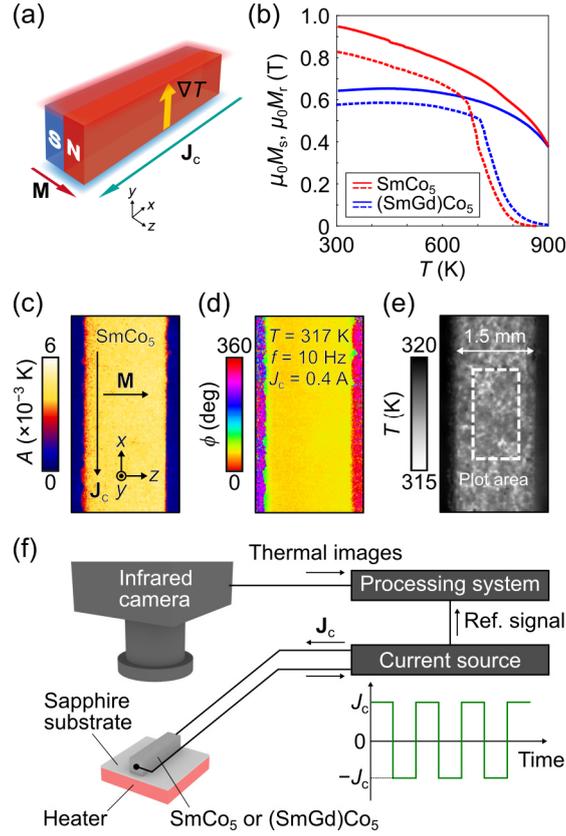

**FIG. 1.** (a) Schematic illustration of AEE in a permanent magnet in the in-plane magnetized configuration. $\mathbf{J}_c$, $\nabla T$, and $\mathbf{M}$ denote the charge current applied to the magnet, temperature gradient generated as a result of AEE, and direction of the remanent magnetization, respectively. (b) Temperature $T$ dependence of the saturation magnetization $\mu_0 M_s$ (solid lines) and the remanent magnetization $\mu_0 M_r$ (dotted lines) for the SmCo$_5$ and (SmGd)Co$_5$ slabs, measured by the vibrating sample magnetometry. $\mu_0$ is the vacuum permeability. (c) and (d) Lock-in amplitude $A$ and phase $\phi$ images for the SmCo$_5$ slab at $T = 317$ K, $f = 10$ Hz, and $J_c = 0.4$ A. $f$ and $J_c$ denote the frequency and amplitude of the square-wave-modulated charge current, respectively. (e) Steady-state thermal image for the SmCo$_5$ slab with the black-ink coating. (f) Schematic illustration of AEE measurements based on LIT. The measurements were performed in the absence of an external magnetic field. The heater and resistance temperature sensor embedded in the Cu stage were used for controlling the sample temperature.

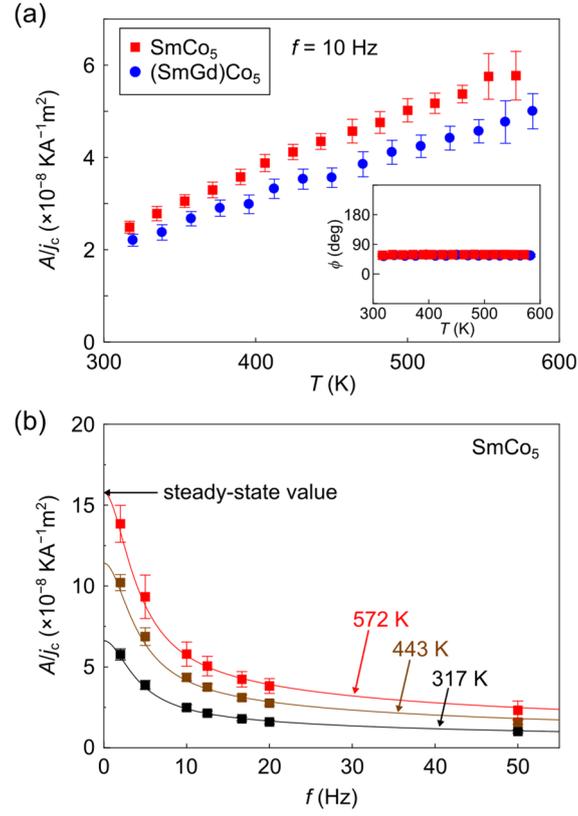

**FIG. 2.** (a) $T$ dependence of $A/j_c$ for the SmCo$_5$ (red squares) and (SmGd)Co$_5$ (blue circles) slabs at $f = 10$ Hz. $j_c$ denotes the amplitude of the square-wave-modulated charge current density. The inset shows the $T$ dependence of $\phi$. The error bars represent the standard deviation of the $A/j_c$ values on the area defined by the white rectangle with a size of $100 \times 40$ pixels in Fig. 1(e). (b) $f$ dependence of $A/j_c$ for the SmCo$_5$ slab at $T = 317$ K (black), 443 K (brown), and 572 K (red). The solid lines show the calculated $f$ dependence of $A/j_c$ for the SmCo$_5$ slab, obtained by solving the one-dimensional heat equation. The calculation procedures are shown in Ref. 8.

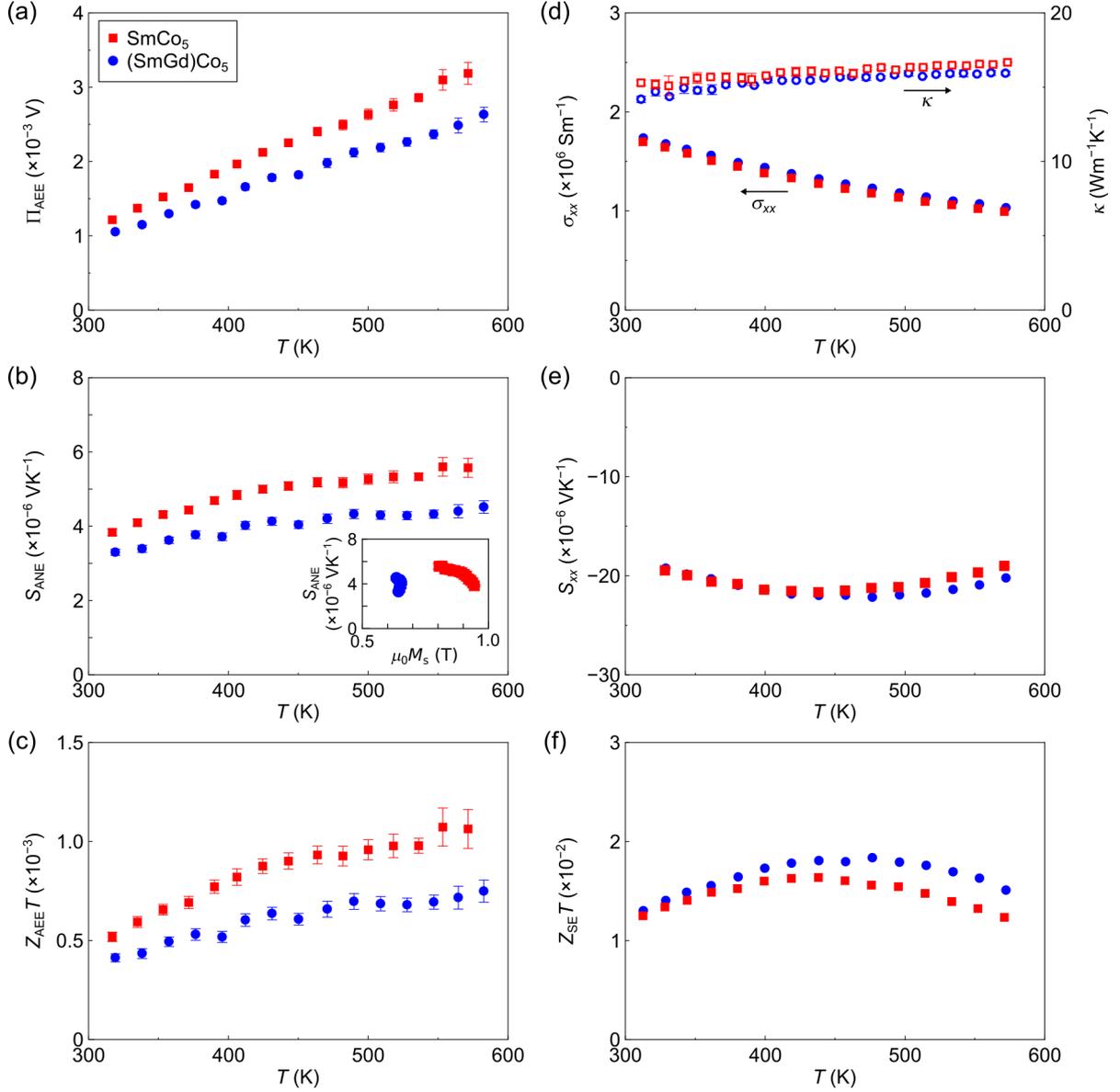

**FIG. 3.** (a) and (b) $T$ dependence of the anomalous Ettingshausen coefficient $\Pi_{AEE}$ and the anomalous Nernst coefficient $S_{ANE}$ (= $\Pi_{AEE}/T$) for SmCo$_5$ (red squares) and (SmGd)Co$_5$ (blue circles). The inset to (b) shows the $\mu_0 M_s$ dependence of $S_{ANE}$. (c) $T$ dependence of the dimensionless figure of merit $Z_{AEE}T$ for AEE. (d) $T$ dependence of the longitudinal electrical conductivity $\sigma_{xx}$ (closed) and the thermal conductivity $\kappa$ (open). (e) and (f) $T$ dependence of the Seebeck coefficient $S_{xx}$ and the dimensionless figure of merit $Z_{SE}T$ for the Seebeck effect.

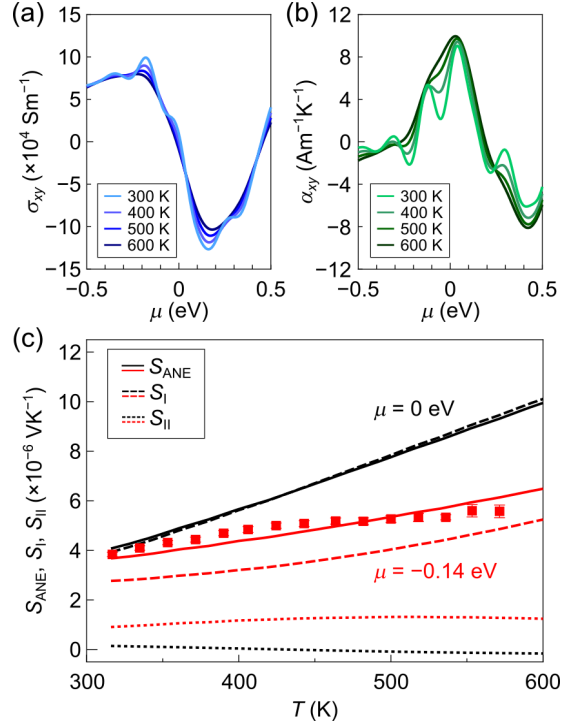

**FIG. 4.** (a) and (b) Chemical potential $\mu$ dependence of the anomalous Hall conductivity $\sigma_{xy}$ and the transverse thermoelectric conductivity $\alpha_{xy}$ of SmCo$_5$ for various values of $T$, obtained from the first-principles calculations. The calculation procedures are shown in Ref. 8. (c) Comparison of the $T$ dependence of the experimentally observed $S_{\mathrm{ANE}}$ values (red squares) with that of $S_{\mathrm{ANE}}$ (solid lines), $S_{\mathrm{I}}$ (dashed lines), and $S_{\mathrm{II}}$ (dotted lines) estimated based on the first-principles calculations for SmCo$_5$. The black (red) lines show the calculated results for $\mu$ = 0 eV (–0.14 eV).